\documentclass[%
 reprint,
 amsmath,amssymb,
 aps,
]{revtex4-2}

\usepackage{graphicx}
\usepackage{dcolumn}
\usepackage{bm}
\usepackage{microtype}
\usepackage{xcolor}

\begin{document}

\preprint{APS/123-QED}

\title{Isotope shifts and hyperfine splitting of the \({}^1S_0 \rightarrow {}^3P_1\) transition in zinc}

\author{Felix Waldherr}
 \email{fwaldher@uni-bonn.de}
\author{Lukas Möller}
\author{Simon Stellmer}%
\affiliation{%
 University of Bonn, Nussallee 12, 53115 Bonn, Germany
}%

\date{\today}

\begin{abstract}
We report laser-induced-fluorescence spectroscopy of the
\({}^1S_0 \rightarrow {}^3P_1\)
intercombination transition in neutral zinc at \(307.6~\mathrm{nm}\).
Isotope shifts are measured for all stable isotopes with kHz-level precision, improving previous
data by about two orders of magnitude. \textcolor{black}{For \(^{67}\mathrm{Zn}\), we resolve
the excited-state hyperfine structure and determine
\(\delta\nu^{67,64}_{\rm COG}=1085.933(7)~\mathrm{MHz}\),
\(A=608.922(3)~\mathrm{MHz}\), and \(B=-18.995(10)~\mathrm{MHz}\).
A King plot comparison with the \({}^1S_0 \rightarrow {}^1P_1\) 214-nm transition results in field- and mass-shift parameters of \(F_{307.6,214}=1.17(5)\) and
\(K_{307.6,214}=-153(60)~\mathrm{GHz\ u}\).} These results provide the spectroscopic
basis for narrow-line cooling and precision measurements based on zinc, including the development of an optical clock.
\end{abstract}

\maketitle


\section{\label{sec:introduction}Introduction}

Precision spectroscopy of atoms and molecules provides accurate frequency
references and sensitive probes of atomic structure and provides the basis for
tests of fundamental physics. A particularly successful class of systems is formed by alkaline-earth and
alkaline-earth-like atoms, \textit{i.e.}, two-valence-electron atoms with a
closed-shell \({}^{1}S_{0}\) ground state. \textcolor{black}{This includes the group-II
alkaline-earth atoms Mg, Ca, Sr, Be, Ra, and Ba, as well as related
alkaline-earth-like systems such as Zn, Cd, Hg, and Yb.} Their level structure supports a diverse set of transitions:
a broad singlet transition provides efficient first-stage cooling, while the narrow
intercombination line \textcolor{black}{enables} second-stage cooling and high-resolution
spectroscopy, ultimately leading to ultra-narrow clock transitions~\cite{witkowski2019,gu2025,miyake2019,tarallo2014}.

The isotope structure of these systems provides an additional degree of
control. Bosonic isotopes with nuclear spin \(I=0\) have hyperfine-free
spectra, while fermionic isotopes exhibit hyperfine structure and a weakly allowed
${}^{1}S_{0}\rightarrow{}^{3}P_{0}$ clock transition through hyperfine
mixing~\cite{aeppli2024}.
\textcolor{black}{The combination} of simple bosonic spectra and richer fermionic structure is
useful for experiments with laser-cooled atoms.

\textcolor{black}{Neutral zinc is an attractive but less explored member of this
alkaline-earth-like class. The $^{1}S_{0}\to{}^{3}P_{0}$ clock transition is predicted to exhibit a small blackbody-radiation shift, reducing sensitivity to one of the important systematic effects in lattice clocks. In addition, the short transition wavelengths of Zn are attractive for experiments that benefit from high spatial resolution.} Laser cooling and trapping  of zinc on the
\({}^{1}S_0\rightarrow{}^{1}P_1\) transition at
\(214~\mathrm{nm}\) have recently been demonstrated
\cite{Roser2024,moeller2025mot}. Second-stage cooling of zinc requires isotope-resolved transition frequencies on the \(307.6~\mathrm{nm}\) intercombination line. This motivates a precise
measurement of the isotope shifts and of the \(^{67}\mathrm{Zn}\) hyperfine
structure. 

Here, we present high-resolution laser-induced-fluorescence spectroscopy of
the \(307.6~\mathrm{nm}\) intercombination line in a thermal beam of neutral
zinc atoms. We determine isotope shifts for the stable isotopes relative to
\(^{64}\mathrm{Zn}\) and resolve the excited-state hyperfine structure of
\(^{67}\mathrm{Zn}\). These measurements provide the frequency references
needed for isotope-selective narrow-line cooling of zinc.

\section{\label{sec:exp_setup}Experimental setup}
\begin{figure*}[!t]
    \centering
    \includegraphics[width = \linewidth]{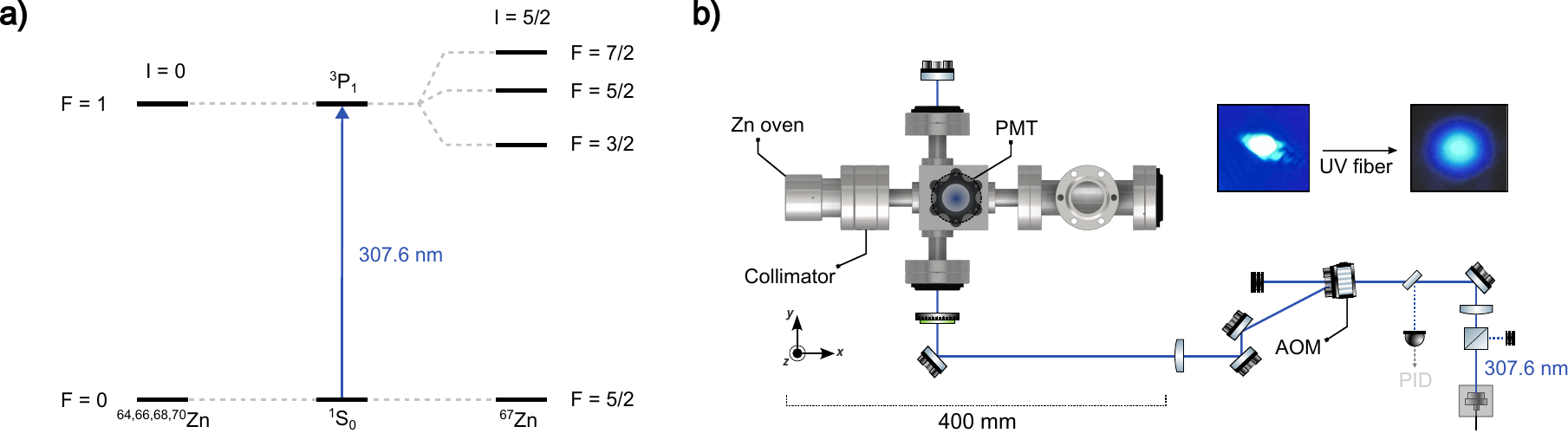}
    \caption{\label{fig:experimental_setup}
    Level structure and experimental setup for spectroscopy of the zinc
    intercombination line. 
    (a) Relevant levels for the
    \({}^1S_0 \rightarrow {}^3P_1\) transition at
    \(307.6~\mathrm{nm}\), showing the hyperfine structure of
    \(^{67}\mathrm{Zn}\).
    \textcolor{black}{(b) Schematic of the atomic-beam fluorescence setup. Zn atoms from a thermal
    oven are collimated by a microtube array and interrogated by a retroreflected
    \(307.6~\mathrm{nm}\) spectroscopy beam, which is mode-cleaned by a photonic crystal fiber (shown in the inset). Fluorescence is detected along the
    $z$-direction with a photomultiplier tube (PMT).}
    }
    \label{fig:experimental_Setup_and_level_scheme}
\end{figure*}

Figure~\ref{fig:experimental_Setup_and_level_scheme} shows a sketch of the
experimental setup.
A thermal beam of neutral Zn atoms is produced in a resistively heated oven
operated at temperatures up to \(400^\circ\mathrm{C}\). \textcolor{black}{The atomic beam defines the $x$-axis and is collimated before entering the spectroscopy chamber by a 22-$\mathrm{mm}$-long stainless-steel microtube array with 150-$\mu \mathrm{m}$-diameter channels, integrated into a heated blind flange. This reduces the transverse velocity spread of the beam.} From a Voigt fit to the Doppler-broadened
spectrum, we obtain a transverse Doppler FWHM of
\(\Delta\nu_{\rm D}=155.2(5)~\mathrm{MHz}\), corresponding to
\(\Delta v_\perp=\lambda\Delta\nu_{\rm D}=47.7(2)~\mathrm{m\ s^{-1}}\).
Interpreting this width as that of a one-dimensional Maxwellian velocity
distribution gives an effective transverse temperature
\begin{equation}
T_\perp =
\frac{m}{k_{\rm B}}
\left(
\frac{\Delta v_\perp}{2\sqrt{2\ln 2}}
\right)^2
\simeq 3.2~\mathrm{K}.
\end{equation}
For an oven temperature of \(400^\circ\mathrm{C}\), the thermal mean velocity
is \(\bar v=\sqrt{8k_{\rm B}T/(\pi m)}\simeq466~\mathrm{m\ s^{-1}}\). The
measured transverse velocity width therefore corresponds to a full angular
divergence of
\(\Delta\theta\simeq2\arctan[\Delta v_\perp/(2\bar v)]=5.9(1)^\circ\). \textcolor{black}{We have not directly quantified the atomic flux. Based on ovens of similar design, we estimate the atomic flux density in the interaction region, located approximately $80\ \mathrm{mm}$ downstream of the final oven collimation aperture, to be on the order of $10^{14}\ \mathrm{atoms\ s^{-1}cm^{-2}}$.}

The spectroscopy light at \(307.6~\mathrm{nm}\) is generated by a
commercial frequency-quadrupled diode-laser system (TOPTICA TA-FHG pro). The fundamental light
near \(1232~\mathrm{nm}\) is stabilized to a home-built
\(L=10~\mathrm{cm}\) reference cavity using the Pound--Drever--Hall locking
scheme~\cite{drever1983}. From a fitted cavity ring-down time of
\(\tau_\mathrm{cav}=22.131(7)~\mu\mathrm{s}\), we infer a cavity linewidth of
\(\Delta\nu_\mathrm{cav}=1/(2\pi\tau_\mathrm{cav})=7.191(2)~\mathrm{kHz}\).
Together with the measured free spectral range, this corresponds to a finesse
of \(\mathcal{F}=208\,600(700)\). The laser frequency is monitored with a
wavelength meter (HighFinesse WS8-10) at the second harmonic near \(616~\mathrm{nm}\).

To improve the mode quality of the laser beam, the UV light is delivered to the experiment through a hydrogen-loaded
photonic-crystal fiber (LMA-10 UV)~\cite{Colombe2014}. We typically obtain a
fiber-coupling efficiency of about \(60\%\), allowing up to \(8~\mathrm{mW}\)
of \(307.6~\mathrm{nm}\) light to be delivered to the apparatus. For the
measurements reported here, we use \(P\simeq1.5~\mathrm{mW}\) in a beam with
a \(1/e^2\) diameter of approximately \(6~\mathrm{mm}\). With
\(I_\mathrm{sat}=\pi h c/(3\lambda^3\tau)\simeq
2.7\times10^{-2}~\mathrm{mW\ cm^{-2}}\), this gives a saturation parameter
of \(s_0=I_0/I_\mathrm{sat}\simeq3.9\times10^2\), so that the transition is
driven deeply into saturation.

After the fiber, the laser polarization is cleaned using a polarizing-beam-splitter cube and
set with a half-wave plate. A small fraction of the polarization-cleaned beam
is picked off with a wedged window and monitored on a photodiode for
intensity stabilization. The spectroscopy beam then propagates along the
\(y\)-axis and intersects the atomic beam at right angles. The perpendicular
alignment is first set using irises before and after the chamber and then
refined by minimizing the offset between the Doppler-free feature and the
center of the Doppler-broadened fluorescence envelope. To generate the
Doppler-free signal, the beam is retroreflected after the vacuum chamber; the
retroreflected light can be coupled back into the photonic-crystal fiber to
verify spatial overlap.

\textcolor{black}{For the spectroscopy scans, the frequency of the cavity-stabilized fundamental light is tuned by adjusting a frequency offset between the laser and the mode of the cavity, using an electro-optic modulator. The scan sequence and RF offsets were controlled by an ARTIQ control system, which in turn is referenced to a Rb atomic clock. The RF frequencies can be set with sub-kHz precision, which is well below the statistical and \textcolor{black}{systematic uncertainties in our measurements}. The wavelength meter with an accuracy of about 2\,MHz is used to monitor the absolute optical frequency and to identify the correct cavity mode, but it is not used for the spectroscopy scans. All frequency intervals, linewidths, scan ranges, and scan steps reported in this work are quoted on the ultraviolet frequency scale. The isotope-shift spectra were recorded with a step size of 40\,kHz.}

Magnetic fields can be applied along the \(x\) and \(y\) axes using two
independent pairs of coils. In the measurements discussed here, the relevant
field is applied along \(x\), parallel to the atomic beam, and defines the
magnetic quantization axis. The spectroscopy light is linearly polarized
parallel to this axis to drive predominantly \(\pi\) transitions. This
polarization also favors fluorescence collection along the vertical \(z\)
direction, since the dipole-emission pattern for \(\pi\)-polarized excitation
has maximum emission perpendicular to the quantization axis.

Fluorescence is collected along \(z\) with a photomultiplier tube
(PMT, Hamamatsu H9306). To suppress scattered light at the detector, the
spectroscopy beam intersects the atomic beam approximately \(2~\mathrm{cm}\)
upstream of the PMT collection volume. At
\(\bar v\simeq466~\mathrm{m\ s^{-1}}\), atoms traverse this distance in
about \(43~\mu\mathrm{s}\). With an excited-state lifetime of
\(\tau\simeq26~\mu\mathrm{s}\), roughly \(19\%\) of the excited atoms remain
in the excited state until they reach the detection region. This spatial
offset preserves a measurable fluorescence signal while efficiently rejecting
prompt scattered light from the excitation region.
For improved detection sensitivity, the spectroscopy light is
amplitude-modulated with an acousto-optic modulator operated at a center
frequency of \(f_\mathrm{c}=200~\mathrm{MHz}\). The modulation frequency is
\(f_\mathrm{mod}=10~\mathrm{kHz}\), and the PMT signal is demodulated with a
lock-in amplifier (MFLI, Zurich Instruments). With this detection scheme, even the weakest naturally
abundant bosonic isotope, \(^{70}\mathrm{Zn}\), is detected with a
signal-to-noise ratio exceeding 10 after \(10~\mathrm{s}\) of averaging per
frequency point.

\section{\label{sec:results}Results} 
Figure~\ref{fig:full_spectrum} shows the Doppler-broadened fluorescence
spectrum of the
\({}^{1}S_{0}\rightarrow{}^{3}P_{1}\)
transition at \(307.6~\mathrm{nm}\). On this scale, the observed line shape
is dominated by the transverse velocity distribution of the thermal atomic
beam. 
\textcolor{black}{The relative amplitudes of the resolved components are consistent with the
natural isotopic abundances listed in Table~\ref{tab:zn_abundances},
allowing the individual isotopes to be assigned in the Doppler-broadened
spectrum.}

\begin{table}[!h]
\caption{\label{tab:zn_abundances}
Natural isotopic abundances of zinc \cite{rosman1998}.
}
\begin{ruledtabular}
\begin{tabular}{ccc}
Isotope & Natural abundance (\%) & Nuclear spin \\ \hline
$^{64}$Zn & $48.63(60)$ & 0 \\
$^{66}$Zn & $27.90(27)$ & 0\\
$^{67}$Zn & $4.10(13)$ & $5/2$\\
$^{68}$Zn & $18.75(51)$ & 0\\
$^{70}$Zn & $0.62(3)$ & 0\\
\end{tabular}
\end{ruledtabular}
\end{table}

\begin{figure}[!b]
    \centering
    \includegraphics[width = \linewidth]{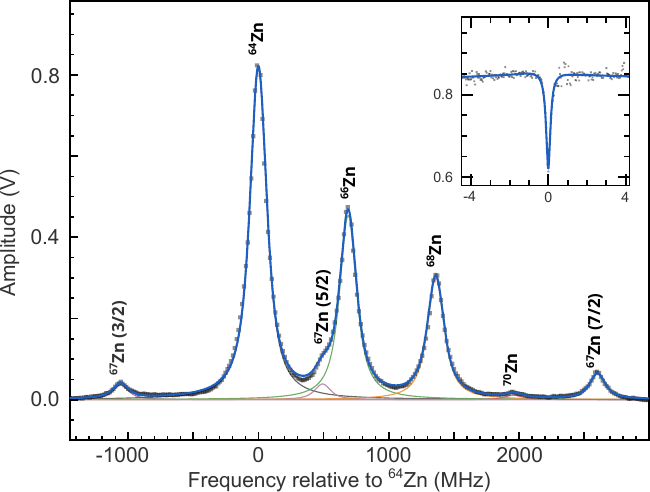}
    \caption{\label{fig:full_spectrum}
    Isotope-resolved fluorescence spectrum of the
    ${}^{1}S_{0}\rightarrow{}^{3}P_{1}$ transition in Zn at 307.6~nm.
    The main panel shows the Doppler-broadened spectrum, with the frequency axis
    referenced to the fitted ${}^{64}\mathrm{Zn}$ resonance. The blue curve is a
    multi-Voigt fit including the bosonic isotopes and the three resolved
    ${}^{67}\mathrm{Zn}$ hyperfine components, while the thin colored curves show
    the individual contributions. The inset shows a Doppler-free spectrum of the
    ${}^{64}\mathrm{Zn}$ resonance used to determine the zero of the frequency
    axis.
    }
    \label{fig:full_spectrum}
\end{figure}

For linearly polarized excitation, the detected fluorescence depends
on the dipole emission pattern. For an even-mass isotope \(e\), we follow Brown \textit{et al.} \cite{brown2013} to model the
fluorescence signal as a modified Voigt signal
\begin{equation}
S^{(e)}(\nu)
=
A_e
V(\nu-\nu_e;\sigma_{\rm D},\gamma)
\left[
1-P_2(\cos\theta)g(\theta_C)
\right],
\label{eq:boson_broad_voigt}
\end{equation}
where \(A_e\) absorbs the isotope abundance, transition strength, and overall
detection efficiency. The angular factor is given by
\begin{equation}
P_2(\cos\theta)=\frac{3\cos^2\theta-1}{2},
\end{equation}
\begin{equation}
g(\theta_C)=\cos\theta_C\cos^2\left(\frac{\theta_C}{2}\right),
\end{equation}
where \(\theta\) is the angle between the excitation polarization and the
detection direction, and \(\theta_C\) is the half-angle of the finite
collection cone. In our setup, a \textcolor{black}{$2$-$\mathrm{inch}$} collection lens is
located about \(60~\mathrm{mm}\) from the fluorescence region, giving
\(\theta_C\simeq23^\circ\). The PMT observes
fluorescence perpendicular to the laser polarization, so that
\(\theta=\pi/2\). This
geometry maximizes the detected fluorescence.

\subsection{\label{subsec:bosonic}Bosonic isotopes}

The \textcolor{black}{even-mass-number isotopes} \(^{64}\mathrm{Zn}\), \(^{66}\mathrm{Zn}\),
\(^{68}\mathrm{Zn}\), and \(^{70}\mathrm{Zn}\) have nuclear spin \(I=0\)
and no hyperfine structure. In the applied magnetic field of \(B\simeq340~\mu\mathrm{T}\), corresponding
to a coil current of 2\,A, the \(\sigma^\pm\)
\((\Delta m_J=\pm1)\) components are shifted away from the central
\(\pi\) \((\Delta m_J=0)\) transition. With the spectroscopy polarization parallel to
the field, we predominantly probe the \(\pi\)-transition, so
each bosonic isotope contributes one fitted resonance for the isotope-shift
analysis.

\begin{figure*}[!t]
    \centering
    \includegraphics[width=\linewidth]{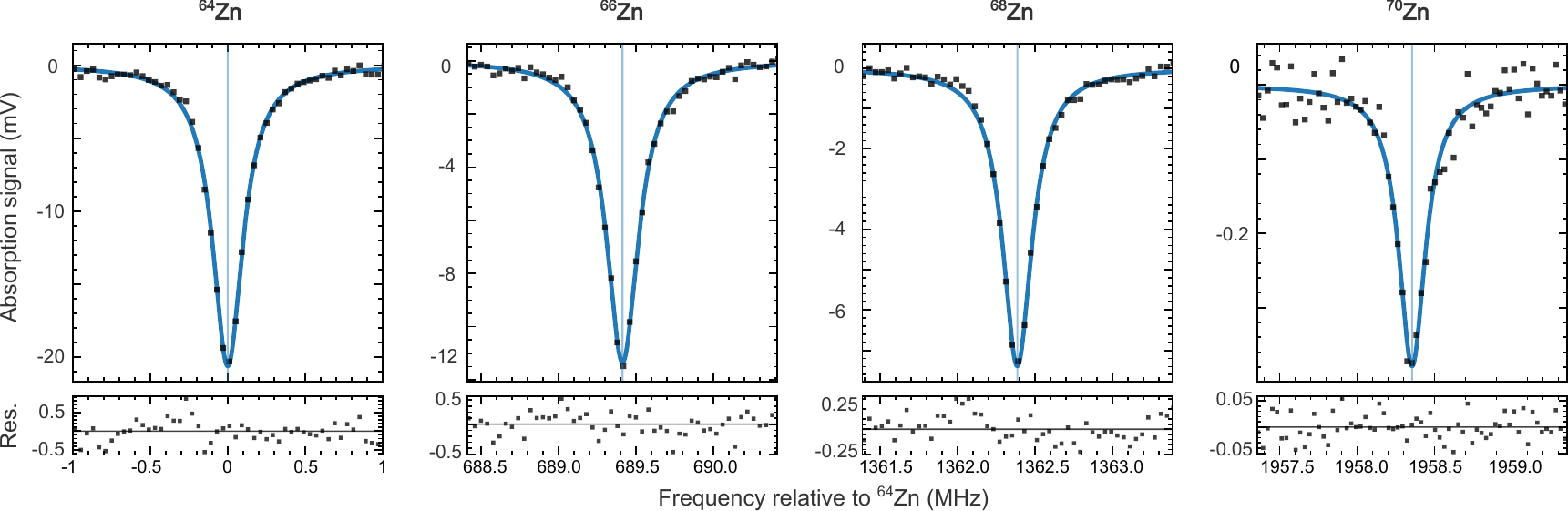}
    \caption{
    Doppler-free spectra of the bosonic zinc isotopes on the
    \({}^1S_0 \rightarrow {}^3P_1\) transition at
    \(307.6~\mathrm{nm}\). Each panel shows the background-subtracted
    fluorescence signal, with the frequency axis referenced to the
    \(^{64}\mathrm{Zn}\) resonance. Black points are the measured data and
    blue curves are Lorentzian fits to the saturation dips. The lower panels
    show the corresponding fit residuals. The fitted centers are used to
    determine the isotope shifts of \(^{66}\mathrm{Zn}\),
    \(^{68}\mathrm{Zn}\), and \(^{70}\mathrm{Zn}\) relative to
    \(^{64}\mathrm{Zn}\).
    }
    \label{fig:bosonic}
\end{figure*}

With the spectroscopy beam retroreflected, atoms with near-zero velocity along
the laser axis interact with both counterpropagating beams, producing a
Doppler-free saturation dip on top of the Doppler-broadened fluorescence
background. The background-subtracted spectra of the bosonic isotopes are
shown in Fig.~\ref{fig:bosonic}.

The local spectrum is fitted with
\begin{equation}
S_{\rm DF}^{(e)}(\nu)
=
S^{(e)}(\nu)
-
\frac{A_{\rm DF}}
{1+4(\nu-\nu_0)^2/\Gamma_{\rm DF}^2},
\label{eq:boson_df_fit}
\end{equation}
where \(S^{(e)}(\nu)\) is the Voigt background from
Eq.~\eqref{eq:boson_broad_voigt}. The Lorentzian term describes the
Doppler-free dip with center frequency \(\nu_0\), FWHM \(\Gamma_{\rm DF}\), and
amplitude \(A_{\rm DF}\). The fixed angular-emission factor is absorbed into
\(A_{\rm DF}\).

\subsection{\label{subsec:fermionic_zn67}Hyperfine structure of $^{67}$Zn}

\begin{figure*}[!t]
    \centering
    \includegraphics[width=\linewidth]{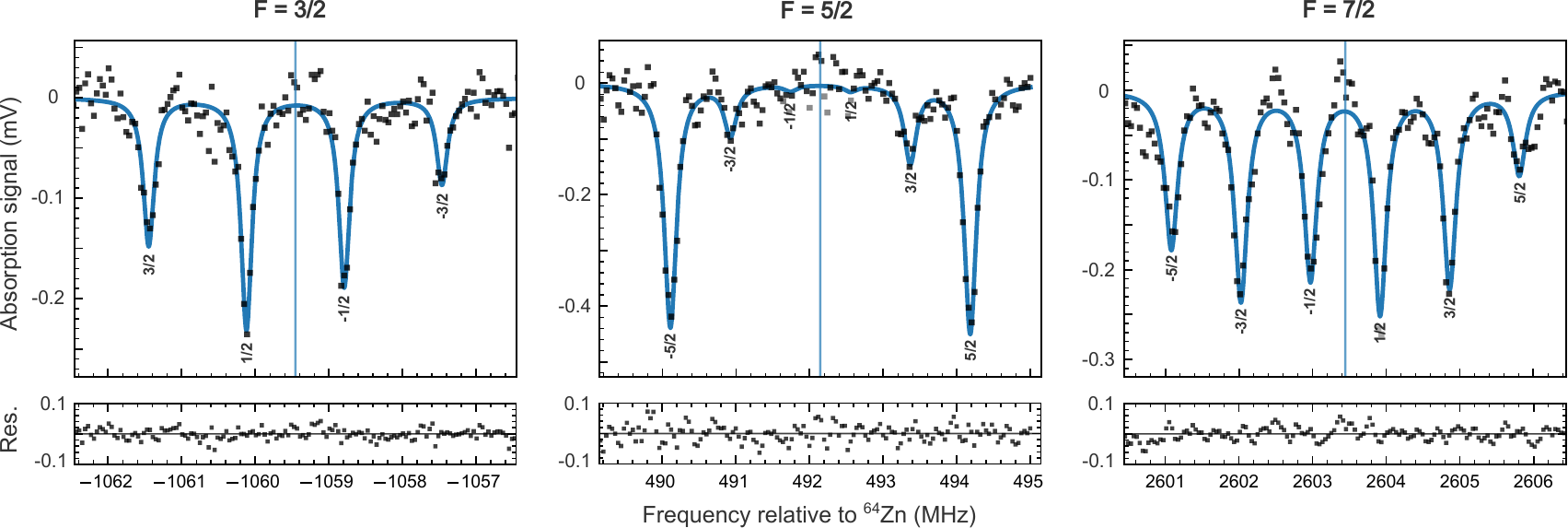}
    \caption{
    Doppler-free spectra of the three resolved hyperfine manifolds of
    $^{67}$Zn. The frequency axis is
    referenced to the $^{64}$Zn resonance. Blue markers indicate the
    fitted line positions used to determine the hyperfine intervals and
    the degeneracy-weighted center-of-gravity isotope shift.
    }
    \label{fig:fermionic_lines}
\end{figure*}

For \(^{67}\mathrm{Zn}\), the Doppler-free spectra resolve all three
excited-state hyperfine manifolds of the \(4s4p{}^{3}P_{1}\) state,
\(F'=3/2,5/2,\) and \(7/2\), as shown in
Fig.~\ref{fig:fermionic_lines}. Under a magnetic field of \(B\simeq340~\mu\mathrm{T}\) applied along the $x$-axis, the Zeeman structure is resolved as well.
The center-of-gravity (COG) isotope shift is obtained from the
degeneracy-weighted average over the three hyperfine manifolds,
\begin{equation}
\nu_{\rm COG}
=
\frac{\sum_{F'}(2F'+1)\nu_{F'}}
     {\sum_{F'}(2F'+1)}
=
\frac{4\nu_{3/2}+6\nu_{5/2}+8\nu_{7/2}}{18}.
\end{equation}
The observed hyperfine-component frequencies can be written as
\begin{equation}
\nu_{F'}=\nu_{\rm COG}+\Delta\nu_{F'},
\end{equation}
where \(\nu_{\rm COG}\) is the fine-structure center-of-gravity frequency
and \(\Delta\nu_{F'}\) are the hyperfine shifts. The latter are given by the
Casimir formula
\begin{equation}
\Delta\nu_{F'}
=
\frac{1}{2}AK
+
B
\frac{\frac{3}{4}K(K+1)-I(I+1)J(J+1)}
     {2I(2I-1)J(2J-1)},
\end{equation}
where
\begin{equation}
K=F'(F'+1)-I(I+1)-J(J+1).
\end{equation}
For \(^{67}\mathrm{Zn}\), the nuclear spin is \(I=5/2\). In the excited
\({}^{3}P_{1}\) state, the electronic angular momentum is
\(J=1\). Taking differences between
the Casimir shifts gives
\begin{equation}
\nu_{5/2-3/2}
=
\Delta\nu_{5/2}-\Delta\nu_{3/2}
=
\frac{5}{2}A-\frac{3}{2}B,
\end{equation}
\begin{equation}
\nu_{7/2-5/2}
=
\Delta\nu_{7/2}-\Delta\nu_{5/2}
=
\frac{7}{2}A+\frac{21}{20}B.
\end{equation}
The extracted intervals, COG shifts, and hyperfine constants A and B are listed in
Table~\ref{tab:zn67_hfs_summary}.
\begin{table}[!h]
\caption{\label{tab:zn67_hfs_summary}
\textcolor{black}{Hyperfine intervals and derived hyperfine constants of the $^{67}$Zn $4s4p{}^{3}P_{1}$ excited state, together with center-of-gravity (COG) isotope shifts for the ${}^{1}S_{0} \rightarrow {}^{3}P_{1}$ transition at 307.6 nm. Uncertainties in parentheses include the statistical fit uncertainty and a conservative 10 kHz second-order Zeeman uncertainty assigned to each individual hyperfine component position, added in quadrature. Further details are found in Section \ref{subsec:uncertainties}.}
}
\begin{ruledtabular}
\begin{tabular}{lcc}
Quantity
& This work
& Previous measurement \\
& MHz
& MHz \\ \hline

\multicolumn{3}{c}{Measured hyperfine intervals} \\ \hline

$\nu_{5/2-3/2}$
& \textcolor{black}{$1550.795(16)$}
& $1551.565(4)^{a}$ \\

$\nu_{7/2-5/2}$
& \textcolor{black}{$2111.283(15)$}
& $2111.300(3)^{a}$ \\ \hline

\multicolumn{3}{c}{Extracted hyperfine constants} \\ \hline

$A({}^{3}P_{1})$
& \textcolor{black}{$608.922(3)$}
& $609.086(1)^{a}$ \\

$B({}^{3}P_{1})$
& \textcolor{black}{$-18.995(10)$}
& $-18.782(8)^{a}$ \\ \hline

\multicolumn{3}{c}{Center-of-gravity isotope shifts} \\ \hline

$\delta\nu^{67,64}_{\rm COG}$
& \textcolor{black}{$1085.933(7)$}
& $1085.7(2.0)^{b}$ \\

$\delta\nu^{67,66}_{\rm COG}$
& \textcolor{black}{$396.522(7)$}
& $397.1(2.0)^{b}$ 

\end{tabular}
\end{ruledtabular}

\begin{flushleft}
$^{a}$Optical double-resonance measurement by Byron
\textit{et al.}~\cite{Byron1964}.\\
$^{b}$Atomic-beam measurement by Campbell
\textit{et al.}~\cite{Campbell1997}.
\end{flushleft}
\end{table}
The measured hyperfine constants also provide useful benchmarks for
atomic-structure theory \cite{sahoo2023}. Multiconfiguration Hartree--Fock (MCHF) and multiconfiguration Dirac--Hartree--Fock
(MCDHF) calculations of the \(4s4p\,{}^{3}P_{1,2}\) states have been used
to extract the \(^{67}\mathrm{Zn}\) nuclear quadrupole moment from
experimental hyperfine constants and calculated electric-field gradients
\cite{bieron2018}. Because the
magnetic-dipole and electric-quadrupole constants are strongly correlated,
small changes in the fitted hyperfine intervals or in the theoretical
treatment of electron correlation, relativistic effects, and spin
polarization can produce appreciable deviations in \(A\) and \(B\). The
present kHz-level values for the $4s4p{}^{3}P_{1}$ state therefore
provide an additional constraint for future many-body calculations of Zn~I \cite{Filippin2017}. \textcolor{black}{We note that our result on the $F=5/2 \rightarrow  F'=3/2$ transition differs significantly from previous work, which used a different spectroscopy scheme \cite{Byron1964}. The reason for this discrepancy is not known.}
\newpage
\subsection{\label{subsec:isotopeshifts}Isotope shift analysis}
Isotope shifts reflect the fact that different isotopes have different nuclear
masses and charge distributions. The mass difference changes the electronic
transition energy through the normal and specific mass shifts, while the change
in nuclear charge radius gives rise to the field shift. The total isotope
shift is therefore commonly written as the sum of a mass-shift and a
field-shift contribution,
\begin{equation}
\delta\nu^{A,64}
=
K\left(\frac{1}{m_A}-\frac{1}{m_{64}}\right)
+
F\delta\langle r^2\rangle^{A,64},
\end{equation}
where \(m_A\) denotes the mass of the isotope with mass number \(A\), \(K\) and \(F\) are the mass- and field-shift constants of
the transition. Table~\ref{tab:isotope_shift_summary} summarizes the
isotope shifts measured for the stable zinc isotopes on the
$^{1}S_{0}\rightarrow{}^{3}P_{1}$ transition. 

\begin{table}[!h]
\caption{\label{tab:isotope_shift_summary}
Summary of the isotope-shift measurements of the zinc
${}^{1}S_{0}\rightarrow{}^{3}P_{1}$ transition at 307.6~nm. For
${}^{67}$Zn, the center-of-gravity (COG) isotope shift is listed.}
\begin{ruledtabular}
\begin{tabular}{lcc}
Quantity
& This work
& Previous measurement \\
& MHz
& MHz \\ \hline

$\delta\nu^{66,64}$
& $689.411(3)$
& $688.6(1.0)^{a}$ \\

$\delta\nu^{67,64}$
& \textcolor{black}{$1085.933(7)$}
& $1085.7(2.2)^{a}$ \\

$\delta\nu^{68,64}$
& $1362.390(3)$
& $1365.2(1.4)^{a}$ \\

$\delta\nu^{70,64}$
& $1958.358(7)$
& $1962.2(1.7)^{a}$ \\ 

\end{tabular}
\end{ruledtabular}
\begin{flushleft}
$^{a}$Atomic-beam measurement by Campbell
\textit{et al.}~\cite{Campbell1997}.
\end{flushleft}

\end{table}

For each isotope pair,
the resonance frequencies were obtained from interleaved measurements of
$^{64}$Zn and the target isotope. This sequence reduces sensitivity to slow
drifts of the experiment, particularly of the reference cavity.
For the bosonic isotopes, the isotope shift is obtained directly from the difference between the fitted
Doppler-free resonance positions. For $^{67}$Zn, the individual hyperfine components were measured separately. The isotope shift of the fermionic isotope is therefore
reported as the center-of-gravity shift. The same data also yield the excited-state hyperfine intervals
and the corresponding magnetic-dipole and electric-quadrupole hyperfine constants.
The residual drift of the experimental setup during each interleaved measurement sequence was determined from the
time dependence of the repeated $^{64}$Zn reference measurements and subtracted from the extracted isotope
shifts. The size of this correction, together with the uncertainty assigned to the remaining uncompensated
drift, is discussed in Sec.~\ref{subsec:uncertainties}. This procedure allows the isotope shifts to be determined
largely independently of the absolute cavity drift over the full measurement time.

To compare the present \(307.6~\mathrm{nm}\) isotope shifts with the previously
measured \({}^{1}S_0\rightarrow{}^{1}P_1\) transition at
\(214~\mathrm{nm}\)~\cite{Roser2024}, we construct a King plot. Isotope
shifts contain both mass-shift and field-shift contributions; comparing two
transitions allows these contributions to be tested for mutual consistency.
For each isotope pair we define the modified isotope shift
\begin{equation}
m\delta\nu^{A,64}
=
\frac{\delta\nu^{A,64}}{\mu^{A,64}},
\qquad
\mu^{A,64}
=
\frac{1}{m_A}
-
\frac{1}{m_{64}} .
\end{equation}
If the mass- and field-shift terms describe the data, the modified
isotope shifts of the two transitions obey a linear relation,
\textcolor{black}{\begin{equation}
m\delta\nu^{A,64}_{307.6}
=
K_{307.6,\,214}
+
F_{307.6,\,214}\,
m\delta\nu^{A,64}_{214},
\end{equation}}
where the slope gives the relative field-shift factor and the intercept gives
the relative mass-shift offset. We fit the \(307.6~\mathrm{nm}\) and
\(214~\mathrm{nm}\) modified isotope shifts using orthogonal distance
regression. The resulting linear fit and its \(1\sigma\) confidence band are
shown in Fig.~\ref{fig:kingplot}. From the
fit we obtain
\[
F_{307.6,\,214}=1.17(5),
\qquad
K_{307.6,\,214}=-153(60)~\mathrm{GHz\ u}.
\]
The near-unity slope reflects the common \({}^{1}S_0\) ground state of both
transitions~\cite{Roser2024}, and the observed linearity shows consistency of
the two isotope-shift data sets within the present uncertainties.

\begin{figure}
    \centering
    \includegraphics[width = \linewidth]{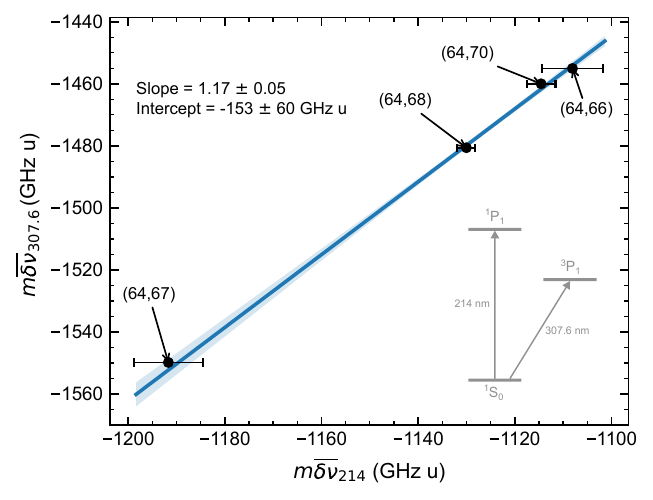}
    \caption{\label{fig:kingplot}
    King plot comparing the modified isotope shifts of the
    \(^{1}S_{0}\rightarrow{}^{3}P_{1}\) transition at 307.6~nm with those of
    the \(^{1}S_{0}\rightarrow{}^{1}P_{1}\) transition at 214~nm \cite{Roser2024}. Each point
    corresponds to one isotope pair relative to \(^{64}\mathrm{Zn}\). The blue line shows a linear fit and
    the shaded band indicates the \(1\sigma\) confidence interval of the fit.
    The inset shows the two electronic transitions used in the comparison.
    }
    \label{fig:kingplot}
\end{figure}

\subsection{\label{subsec:uncertainties}Systematic effects and residual uncertainties}

We evaluate the corrections applied to the measured isotope shifts and the
residual uncertainty from the dominant systematic effects. The resulting
uncertainty budget is summarized in Table~\ref{tab:systematic_budget}.

\begin{figure}[!h]
    \centering
    \includegraphics[width = 1 \linewidth]{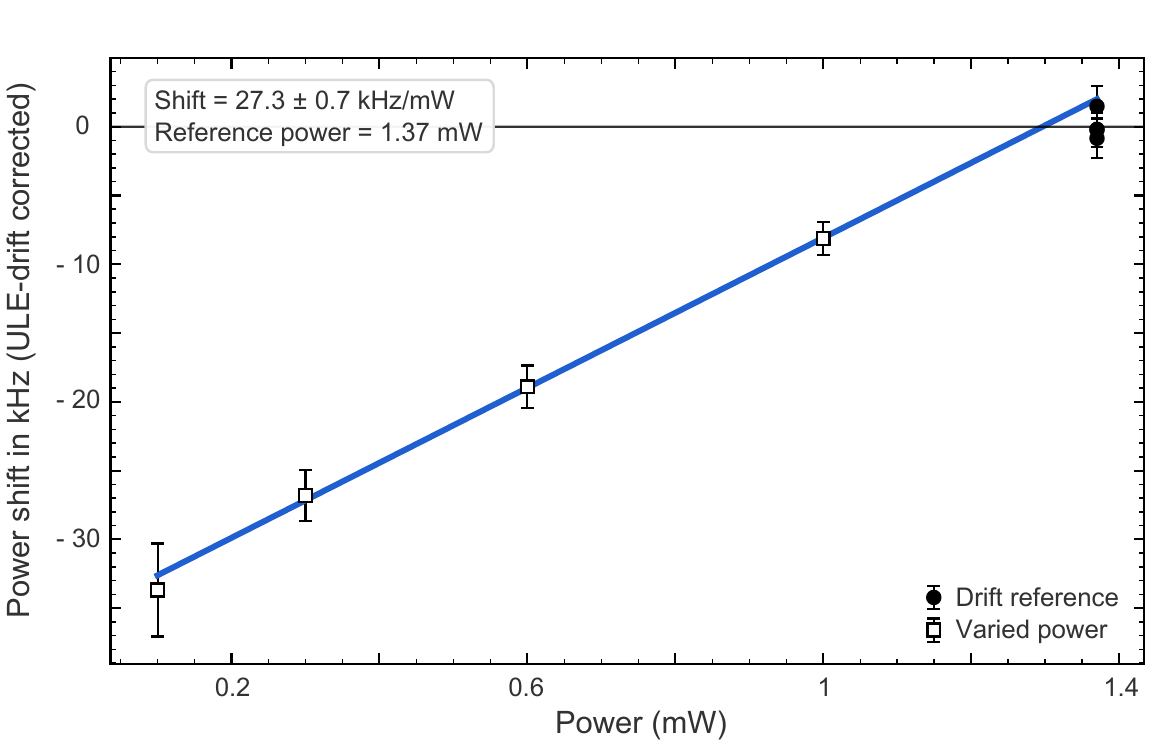}
    \caption{\label{fig:power_shift}
\textcolor{black}{Power-dependent line shift of the ${}^{1}S_{0}\rightarrow{}^{3}P_{1}$
transition frequency at 307.6~nm for $^{64}\text{Zn}$ after correction for the linear reference-cavity
drift. The line centers are plotted relative to the reference power of
1.37~mW.}
}
    \label{fig:int_Dep}
\end{figure}

\begin{table*}[!t]
\caption{\label{tab:systematic_budget}
\textcolor{black}{Uncertainty budget for the isotope-shift measurements of the zinc
${}^{1}S_{0}\rightarrow{}^{3}P_{1}$ transition at 307.6~nm.
The common contributions are applied to all isotope pairs, while the
geometry term is isotope-dependent because it arises from mass-dependent
velocity-class selection. For ${}^{67}$Zn, an additional conservative
second-order Zeeman uncertainty is included in the center-of-gravity
determination, obtained by assigning 10~kHz to each individual hyperfine
component position and propagating this contribution to the COG.
Statistical and systematic contributions are added in quadrature to obtain
the final uncertainties.}}
\begin{ruledtabular}
\begin{tabular}{lcccc}
Contribution
& ${}^{64}$Zn--${}^{66}$Zn
& ${}^{64}$Zn--${}^{67}$Zn 
& ${}^{64}$Zn--${}^{68}$Zn
& ${}^{64}$Zn--${}^{70}$Zn \\
& kHz & kHz & kHz & kHz \\ \hline

\multicolumn{5}{c}{Common systematic contributions} \\ \hline

Reference-cavity drift
& \multicolumn{4}{c}{$1.0$} \\

Intensity dependence
& \multicolumn{4}{c}{$1.9$} \\

Lock-point shifts
& \multicolumn{4}{c}{negligible} \\

Residual $B$-field dependence
& \multicolumn{4}{c}{negligible} \\

Oven-temperature dependence
& \multicolumn{4}{c}{negligible} \\ \hline

Common systematic subtotal
& \multicolumn{4}{c}{$2.1$} \\ \hline

\multicolumn{5}{c}{Isotope-dependent systematic contributions} \\ \hline

Geometry / velocity-class selection
& $1.3$
& $2.0$
& $2.6$
& $3.8$ \\

\textcolor{black}{Second-order Zeeman contribution}
& -- 
& \textcolor{black}{$6.0$}
& --
& -- \\ \hline

Total systematic uncertainty
& $2.5$
& \textcolor{black}{$6.7$}
& $3.4$
& $4.4$ \\

Statistical uncertainty
& $1.5$
& $2.6$
& $0.7$
& $6.0$ \\ \hline

Final uncertainty
& $2.9$
& \textcolor{black}{$7.2$}
& $3.4$
& $7.4$ \\

\end{tabular}
\end{ruledtabular}
\end{table*}
\paragraph{Reference-cavity drift}
The reference cavity operates at the fundamental wavelength, but all drift
rates are quoted on the UV frequency axis after frequency quadrupling. On
this scale, the thermal cavity drift can reach about
\(2~\mathrm{kHz/min}\). Since the isotope shifts are measured in an
interleaved sequence, the common linear drift largely cancels between
neighboring isotope scans. The relevant contribution is therefore the
uncertainty of the fitted drift slope, which we determine to be below
\(0.05~\mathrm{kHz/min}\). For a typical isotope-pair measurement duration of
\(20~\mathrm{min}\), this gives a residual uncertainty of
\(1.0~\mathrm{kHz}\).
\paragraph{Intensity dependence}
\textcolor{black}{Intensity-dependent line-center shifts were characterized by repeating
\(^{64}\mathrm{Zn}\) measurements at different laser powers. The
observed dependence may include AC Stark shifts as well as
intensity-dependent line-shape effects from saturation, optical pumping, or
residual asymmetries of the Doppler-free signal. To isolate the power-dependent shift of the resonance position, we first recorded several spectra at a fixed reference power and used them to determine the residual drift rate at constant power. This drift correction, applied in the same way as for the isotope-shift measurements, removes slow experimental frequency drifts such as those caused by the reference cavity. The shift as a function of power is shown in Fig.~\ref{fig:int_Dep}.}

\textcolor{black}{A linear fit to the power-dependent line shift has a slope of $\beta_P = 27.3(7)~\mathrm{kHz/mW}$. The power stability during the isotope-shift measurements was estimated to be $\Delta P = 50~\mu\mathrm{W}$. Since an isotope shift is obtained from the difference of two independently measured line centers, the corresponding uncertainty is obtained by adding the two power-related uncertainties in quadrature,
$\sigma_P = \sqrt{2}\,\beta_P \Delta P\approx 2~\mathrm{kHz}$. 
The uncertainty of \textcolor{black}{$\beta_P$ is negligible} at this level.}

\paragraph{Instability of the lock point}
For each isotope change, the EOM frequency is changed and the laser is
relocked to the reference cavity. A change of the lock point, for example due
to an electronic offset of the PDH error signal or a slightly different
relocking condition, would therefore appear as an additional frequency offset
between isotope scans. We tested this by repeatedly measuring the
\(^{64}\mathrm{Zn}\) resonance with the cavity lock kept continuously engaged
and after intentionally relocking the laser between successive scans. No
systematic difference between the two data sets is observed within the
statistical scatter of the fitted line centers. We therefore apply no
lock-point correction and bound this contribution to be negligible at the
present level of precision.

\paragraph{Zeeman shift}
Residual magnetic-field effects were tested by repeating the interleaved
\(^{64}\mathrm{Zn}\)--\(^{66}\mathrm{Zn}\) isotope-shift measurement for
different applied \(x\)-field currents. We find a small current dependence of
\(2.1(2)~\mathrm{kHz/A}\). For the measurements, variations in current
between interleaved scans are estimated to be
\(\Delta I\simeq1~\mathrm{mA}\), giving a residual shift of
\(0.002~\mathrm{kHz}\).

\textcolor{black}{For \(^{67}\mathrm{Zn}\), the second-order Zeeman shift was estimated from perturbative mixing of the excited-state hyperfine levels. For the residual magnetic field of approximately 3.4\,G, the calculated quadratic shifts of the individual $m_F$ components are up to $15~\mathrm{kHz}$. Depending on the $m_F$ components and their relative weights, this corresponds to possible shifts of the extracted hyperfine intervals at the level of $10~\mathrm{kHz}$. Since no full magnetic-field-dependent splitting curve was recorded, we do not apply this estimate as an explicit correction, but instead include a conservative second-order Zeeman uncertainty of $10~\mathrm{kHz}$ for the $^{67}\mathrm{Zn}$ hyperfine intervals in the final uncertainty budget.}

\paragraph{Oven temperature}
Possible oven-temperature dependent shifts were tested by repeating the
interleaved \(^{64}\mathrm{Zn}\)--\(^{66}\mathrm{Zn}\) isotope-shift
measurement at \(411^\circ\mathrm{C}\) and \(430^\circ\mathrm{C}\). No change was observed within the experimental uncertainty.

\paragraph{Velocity-class selection}
A residual angular mismatch between the forward and retroreflected
spectroscopy beams can make the Doppler-free signal sensitive to a selected
transverse velocity class rather than to the center of the Doppler-broadened
velocity distribution. We observe that the horizontal alignment of the
retroreflection mirror has a strong influence on the line shape: intentional
misalignment changes the asymmetry and quality of the Doppler-free feature and
shifts the fitted line center. We quantify this effect by deliberately
changing the retroreflection angle with a micrometer screw over a range of
\(\pm3~\mathrm{mrad}\). A linear fit to the measured line-center shifts gives
an alignment sensitivity of approximately
\(
\frac{\partial \nu}{\partial \theta}
\simeq 3.1\times10^{2}~\mathrm{kHz/mrad}.
\)

During normal measurements the retroreflected beam was recoupled into the
photonic-crystal fiber after propagation over approximately \(2~\mathrm{m}\).
The coupling efficiency was stable to within about \(10\%\). For a Gaussian
mode with mode-field diameter \(8.4~\mu\mathrm{m}\), a \(10\%\) coupling loss
corresponds to a transverse mode mismatch of order \(1{-}2~\mu\mathrm{m}\) at
the fiber input. Since the coupling lens partly converts angular deviations
into position at the fiber, we do not use this as an exact pointing
measurement. Instead, we take the full mode-field diameter as a conservative
upper bound on the effective beam displacement, corresponding to
\(
\delta\theta \simeq \frac{8.4~\mu\mathrm{m}}{2~\mathrm{m}}
=4.2~\mu\mathrm{rad}.
\)
Applying the measured alignment sensitivity to this angular scale gives
\(
\sigma_{\mathrm{geom}}^{66,64}
\simeq
\delta\theta \cdot \frac{\partial \nu}{\partial \theta}
\simeq1.3~\mathrm{kHz}.
\)
Since this effect is tied to the selected transverse velocity class, we scale
it with the thermal velocity dependence \(v_\perp\propto m^{-1/2}\),
\begin{equation}
\sigma_{\mathrm{geom}}^{A,64}
=
\sigma_{\mathrm{geom}}^{66,64}
\frac{\left|1-\sqrt{64/A}\right|}
     {\left|1-\sqrt{64/66}\right|},
\end{equation}
using the \(^{64}\mathrm{Zn}\)--\(^{66}\mathrm{Zn}\) pair as the reference.

The total systematic uncertainty for each isotope pair is obtained by adding
the common systematic subtotal and the isotope-dependent geometry term in
quadrature. The final uncertainty is then obtained by adding this systematic
uncertainty and the statistical fit uncertainty in quadrature.

\section{\label{sec:conclusion}Conclusion \& discussion}
We have performed high-resolution spectroscopy of the
$^{1}S_{0}\rightarrow{}^{3}P_{1}$ intercombination line in neutral zinc
using laser-induced fluorescence from a thermal atomic beam. Isotope shifts
for all stable isotopes were measured relative to $^{64}$Zn using
interleaved scan sequences, in which repeated measurements of the reference
isotope were used to correct the drift of the experimental setup. The shifts were measured with
kHz-level precision, improving the accuracy of the available
$307.6~\mathrm{nm}$ isotope-shift data by about two orders of magnitude.

For the bosonic isotopes, the measured shifts are consistent with previous
work while providing substantially smaller uncertainties. For the fermionic
isotope $^{67}$Zn, the resolved $F'=3/2$, $5/2$, and $7/2$ hyperfine
manifolds allow us to determine the center-of-gravity isotope shift and to
extract the magnetic-dipole and electric-quadrupole hyperfine constants of
the $4s4p{}^{3}P_{1}$ state. The observed deviations of the extracted
$A$ and $B$ constants from previous values are small on the MHz scale, but
significant compared with the quoted kHz-level uncertainties, and therefore
deserve further investigation with independent measurements.

The present results provide a precise reference data set for the zinc
intercombination line. This is a useful step toward implementing and
characterizing second-stage cooling on this narrow transition, and more
generally toward improving the control of high-precision zinc experiments. In combination
with isotope shifts on other transitions, the data also enable King plot
consistency checks of mass- and field-shift contributions. A particularly interesting future target is the ultra-narrow
${}^{1}S_{0}\rightarrow{}^{3}P_{0}$ clock transition, which would provide
an additional optical transition for King plot comparisons.

\begin{acknowledgments}
We acknowledge funding from the Deutsche Forschungsgemeinschaft DFG through grants INST 217/978-1 FUGG and 496941189, as well as through the Cluster of Excellence ”ML4Q” (EXC 2004/1 – 390534769) and from the European Commission through project 101080164 ”UVQuanT”. We thank D. Röser for early experimental work, and F. Wolf from PTB for providing us with the UV fiber. We thank the entire team of the UVQuanT consortium for inspiring discussions and technical advice. Data underlying the results presented in this paper are not publicly available at this time, but may be obtained from the authors upon reasonable request.
\end{acknowledgments}


\providecommand{\noopsort}[1]{}\providecommand{\singleletter}[1]{#1}%

\end{document}